\documentclass[twocolumn,showpacs,superscriptaddress,10pt]{revtex4-1} 
\usepackage{amsmath,amssymb,graphicx}
\usepackage{color}
\usepackage{bm}

\newcommand{\rev}[1]{#1}

\begin{document}

\title{Closed-form expressions for nonparaxial accelerating beams with pre-engineered trajectories}

\author{Raluca-Sorina Penciu}
\author{Vassilis Paltoglou}
\author{Nikolaos K. Efremidis} \email{Corresponding author: nefrem@uoc.gr}
\affiliation{Department of Mathematics and Applied Mathematics, University of Crete, 70013 Heraklion, Crete, Greece}

\begin{abstract}
In this letter, we propose a general real-space method for the generation of nonparaxial accelerating beams with arbitrary predefined convex trajectories. Our results lead to closed-form expressions for the required phase at the input plane. 
We present such closed-form results for a variety of caustic curves:  besides circular, elliptic, and parabolic, we find 
for the first time general 
power-law and exponential trajectories. Furthermore, by changing the initial amplitude we can design different intensity profiles along the caustic. 
\end{abstract}



\maketitle

Since 2007, when exponentially truncated Airy beams were predicted~\cite{sivil-ol2007} and observed~\cite{sivil-prl2007}, the study of accelerating waves has attracted a lot of attention (see for example the recent review article~\cite{hu-springer2012}). Such accelerating Airy waves follow parabolic caustic trajectories and were first predicted in the context of quantum mechanics~\cite{berry-ajp1979}. 
Furthermore accelerating beams with trajectories different than parabolic have been found~\cite{green-prl2011,chrem-ol2011,froeh-oe2011}.
The interest in accelerating waves is triggered by several associated applications including filamentation~\cite{polyn-science2009}, particle manipulation~\cite{baumg-np2008,zhang-ol2011}, micromachining~\cite{mathi-apl2012}, and Airy plasmons~\cite{salar-ol2010,zhang-ol2011-plasmon,minov-prl2011}.
In higher dimensions the radial caustic collapse of an accelerating beam generates a focusing effect abrupter as compared to the Lorenzian distribution of a lens focused Gaussian beam~\cite{efrem-ol2010}. Such abruptly autofocusing waves have found applications in generating ablation spots~\cite{papaz-ol2011} as well as in particle manipulation~\cite{zhang-ol2011}. Besides optics, curved and accelerated waves have been considered in different settings such as in electron beams~\cite{voloc-nature2013}, in matter waves~\cite{efrem-pra2013}, and in antenna arrays~\cite{chrem-ieee2013}. In addition to caustic beams, Bessel-like beams can be made to bend along arbitrary trajectories~\cite{chrem-ol2012-bessel,zhao-ol2013}.

In the nonparaxial regime caustic beams offer the advantage of bending at large angles. In this respect, accelerating beams following circular trajectories that can bend up to an almost $90^\circ$ angle have been demonstrated~\cite{courv-ol2012,kamin-prl2012}. Such beams with elliptic and parabolic trajectories have also been found~\cite{zhang-prl2012,aleah-prl2012,bandr-njp2013}. Nonparaxial accelerating beams following arbitrary trajectories have been observed in~\cite{mathi-ol2013} by numerically determining the Fourier space phase profile \rev{and in~\cite{epste-prl2014} using plasmonics}. Nonparaxial Bessel-like beams generated due to the presence of an extended focus, rather than a caustic, following arbitrary trajectories have been proposed~\cite{chrem-pra2013}.

In this work we propose a general real-space method for the generation of nonparaxial caustic accelerating beams that follow arbitrary predefined convex trajectories. More importantly, we present specific classes of trajectories where the phase is computed in closed form. Specifically, we analytically compute the initial phase profile required for nonparaxial accelerating beams following not only circular, elliptic, and parabolic 
(which are described in terms of known functions)
but also for the first time for
general power-law and exponential trajectories. By changing the initial amplitude profile we can generate accelerating beams with an intensity profile that is more uniform along the caustic or more intense in different locations of the trajectory.

Let us consider the propagation of a monochromatic optical wave in a homogeneous isotropic medium with refractive index $n$. The divergence condition $\nabla\cdot\bm E=0$ is satisfied by choosing the electric field to be polarized along the $y$-direction and to depend only on the transverse coordinate $x$ and the propagation coordinate $z$ as $\bm E = \hat{\bm y}\psi(x,z)e^{-i\omega t}$. Then the optical field amplitude satisfies the Helmholtz equation in two dimensions
\begin{equation}
(\nabla^2 +k^2) \psi = 0, \label{Helmholtz}
\end{equation}
where $k=n \omega/c= 2\pi/\lambda$ and $\nabla^2=\partial_x^2+\partial_z^2$. 
Using the Rayleigh-Sommerfeld formulation the dynamics of an arbitrary initial condition $\psi_0(\xi)=A(\xi)e^{i\phi(\xi)}$ is given by
\[
\psi(x,z) = 2\int_{-\infty}^\infty \psi_0(\xi)
\frac{\partial G(x,z;\xi)}{\partial z}d\xi
\]
where $G(x,z;\xi)= -( i/4)H_0^{(1)}(kr)$ is Green's function of the Helmholtz equation,  $H_0^{(1)}$ is a Hankel function, and $r=|\bm r|=\sqrt{(x-\xi)^2+z^2}$. Utilizing the asymptotic form
$
H_0^{(1)}(kr) \sim
\sqrt{2/(\pi k r)}
e^{i(kr-\pi/4)}
$
and applying the stationary phase approximation we obtain the ray equation 
$
x=\xi+r\phi_\xi(\xi)/k
$ 
which after some calculations becomes
\begin{equation} \label{ray_eq}
x = \xi -k'_z(\phi_\xi(\xi))z.
\end{equation}
where 
\begin{equation}
k_z(k_x) = \sqrt{k^2-k_x^2}
\label{kzofkx}
\end{equation}
is the dispersion relation as obtained by selecting the first quadrant from $k_x^2+k_z^2=k^2$.

Applying the implicit function theorem on Eq.~(\ref{ray_eq}) we derive the caustic trajectory of the beam as
$
1 = z k_z''(\phi_{\xi}(\xi)) \phi_{\xi \xi} (\xi).
$
We would like to generate a nonparaxial accelerating beam following a predefined convex trajectory of the form $x_c=f(z_c)$. Noting that the rays should be tangent to the trajectory, we obtain the following relation between the beam trajectory and the initial phase of the beam 
\begin{align} \label{kzp_fp}
k'_z(\phi_\xi (\xi)) = -f'(z_c(\xi)),
\end{align}
where the function $z_c(\xi)$ is calculated from $\xi  = f(z_c) -f'(z_c) z_c$. Utilizing~\eqref{kzp_fp} we can compute $\phi_\xi(\xi)$, from which the initial phase $\phi(\xi)$ is derived by direct integration. This procedure can be implemented either analytically or numerically, depending on the choice of the function $f(z_c)$. In the following paragraphs we are going to apply this method to particular classes of trajectories in which the calculations are carried out analytically leading to closed-form expressions for the phase. 

As a first example, let us consider nonparaxial accelerating beams following circular trajectories~\cite{courv-ol2012,kamin-prl2012}. Rather than choosing a Bessel function as initial condition, our method provides analytic expressions for the phase, allowing us to arbitrarily choose the amplitude function to amplify or suppress the intensity profile along specific parts of the trajectory. Selecting the first quadrant of the circle  
\begin{align} 
x_c^2+z_c^2=R^2 \label{circ_caust_eq}                                                                    \end{align}
results to $x_c = f(z_c) = \sqrt{R^2 -z_c^2}$, where $R$ is the caustic radius. The relation between the transverse coordinate at the input plane and $z_c$ is then given by 
\begin{align} \label{zccirc}
z_c(\xi) = (R/\xi) \sqrt{\xi^2 -R^2}.
\end{align}
Combining Eqs.~\eqref{kzofkx}, \eqref{kzp_fp} and \eqref{zccirc} we find that
$\phi_\xi(\xi) = k_x(\xi) = \pm k \sqrt{\xi^2 - R^2}/\xi$. Finally, integrating the above equation, we obtain the following expression for the initial phase
\begin{align}
\phi(\xi) = k R [ \sqrt{ (  \xi/R )^2 -1} - \sec^{-1} ( \xi/R ) ] .
\end{align}
The initial field amplitude is taken to be $A(\xi) = 0 $ for $0 < \xi < R $ and $A(\xi) = \xi^{-\gamma}$ for $\xi \geq R$, where $\gamma > 0$. 

\begin{figure}[tp]
\centering
\includegraphics[width=\columnwidth]{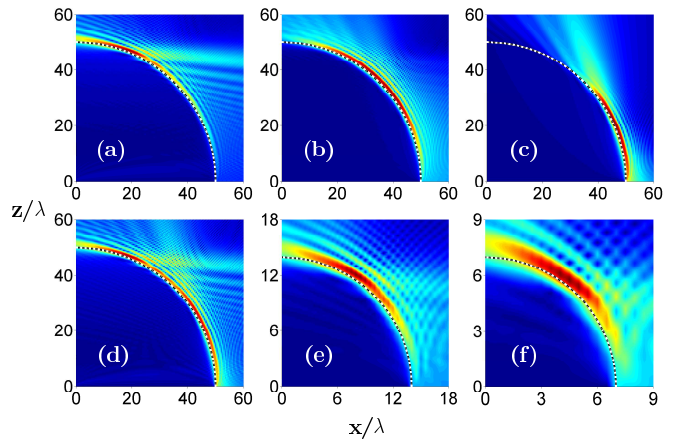} 
 \caption{Amplitude profile of accelerating beams along circular caustic trajectories. In the upper panel $R = 50\lambda$ and (a) $\gamma = 1/4$, (b) $\gamma = 1$, and (c) $\gamma = 5$. 
(d) $R = 50\lambda$ and $A(\xi) = 4/\xi^{10}+1/\xi+1/\xi^{1/10}$, (e) $R=14 \lambda$, $\gamma=1/2$ and (f) $R=7 \lambda$, $\gamma=1/2$.
The white/black dashed line represents the predefined trajectory of the beam.\label{fig1}}
\end{figure}
Typical examples of accelerating beams with circular trajectories are shown in Fig.~\ref{fig1}. The simulation results are obtained by numerically solving Eq.~\eqref{Helmholtz}. Note that in all the figures the coordinates $x$, $z$ are scaled according to the wavelength $\lambda$. The effect of the choice of the initial amplitude function is presented in the top row; 
By changing $\gamma$, we utilize the fact that rays emerging from smaller $\xi$ contribute to earlier stages of the caustic trajectory (smaller values of the $z_c$) making the caustic more prominent at different locations. 
Specifically, in Fig.~\ref{fig1}(a) $\gamma=1/4$ and the caustic is stronger for polar angles close to $\theta=\pi/2$. In Fig.~\ref{fig1}(c), $\gamma=5$ and the caustic is more intense closer to the initial plane. For intermediate values of $\gamma$ the caustic focuses in between as shown in Fig.~\ref{fig1}(b) for $\gamma=1$. By engineering the initial envelope, the caustic remains quasi-invariant to an almost $90^o$ arc as shown Fig.~\ref{fig1}(d). The effect of the size on the beam trajectory is shown in Figs.~\ref{fig1}(e)-(f) where the circle radius is reduced to $R=14\lambda$ and $R=7\lambda$ respectively. We note that, due to the generation of evanescent waves, the beam gets more distorted as the radius becomes smaller.

Let us now consider accelerating beams following elliptic trajectories~\cite{zhang-prl2012,aleah-prl2012,bandr-njp2013} of the form
\begin{align}
x_c = f(z_c) = [ R^2- ( z_c/\alpha )^2 ]^{1/2}.
\label{elip_caust_eq}
\end{align}
Following the above described procedure we analytically computed the initial phase of the beam
\begin{equation}
\phi = k R
[ E ( \sin^{-1} u(\xi) | \tau ) - E ( \sin^{-1} u(R) | \tau ) ], 
\end{equation}
where $u(x)=x/(R\sqrt{\tau})$, $\tau=1-\alpha^2$, and $E(\phi|m)$ is the elliptic integral of the second kind for $\alpha < 1$ and 
\begin{align}
\phi = k R 
[
E (i \sinh^{-1}  u(R) | \tau )  
- 
E (i \sinh ^{-1} u(\xi) | \tau ) 
]
\end{align}
where $u(x)=x/(R\sqrt{-\tau})$, for $\alpha > 1$. 

\begin{figure}[tp]
\centering
\includegraphics[width=\columnwidth]{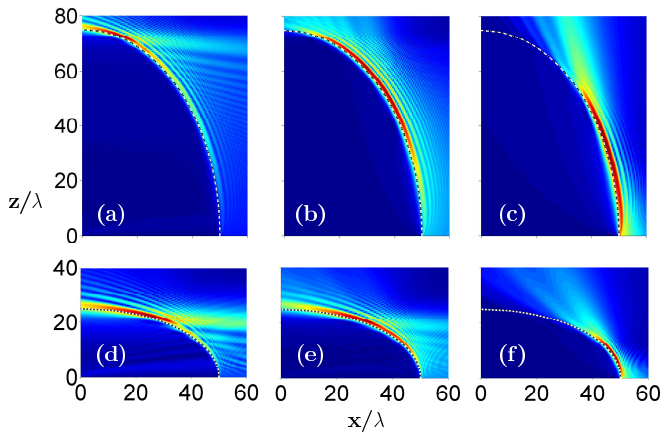} 
\caption{Amplitude profile of accelerating beams following elliptic caustic trajectories for $R=50\lambda$ and $\alpha=3R/2$ (upper panel) and $\alpha = R/2$ (lower panel), (a) $\gamma=1/4$ (b) $\gamma = 1$, (c) $\gamma = 5$,  (d) $\gamma = 1/10$, (e) $\gamma = 1/2$, and (f) $\gamma = 5$. The white/black dashed line represents the pre-defined trajectory of the beam. \label{fig2}} 
\end{figure}

Typical examples of elliptic beam trajectories are shown in Fig.~\ref{fig2}. Specifically, in the upper panel the caustics are constructed from rays launched from the major axis ($\alpha > R$) of the ellipse, while in the lower panel the rays are launched from the minor axis ($\alpha < R$) of the ellipse. The input wave envelope, $A(\xi)$ has the same expression as in the circular trajectory case. By changing the parameters of the amplitude function we generate in the three columns beams that are stronger in the earlier, the intermediate, and the later stages of the caustic, respectively. Such a flexibility might be useful for specific applications, such as micromachining and particle manipulation. 

Let us now consider general power-law trajectories of the form
\begin{align}
x_c = f(z_c) = \alpha z_c ^\beta \label{powereq}
\end{align}
where $\alpha >0$ and $\beta>1$. Following the relevant calculations we find that the input phase profile is given by 
\begin{align}
\phi(\xi) = 
\frac{\alpha \beta^2 k \xi (\alpha\zeta)^{\frac{1}{\beta}-1}}{2 \beta -1}
 {}_2F_1 ( a, b ; c ; -\alpha^{2/\beta} \beta^2\zeta^{\frac{2}{\beta}-2}  ) ,
\end{align}
where ${}_2F_1$ is a hypergeometric function, $\zeta=(\beta-1)/(-\xi)$ $a = 1/2$, $b = 1+ 1/[2(\beta -1)]$, and $c=2+1/[2(\beta-1)]$.  The field amplitude in this case it taken to be $A(\xi) = (\delta - \xi)^{-\gamma}$, where $\delta>0$ for $\xi \leq 0$ and $A(\xi) = 0$ for $\xi > 0$. If we scale the coordinates $(x_c, z_c)$ according to the wavelength, equation \eqref{powereq} becomes 
\begin{align}
x_c/\lambda = \tilde{\alpha} \left( z_c/\lambda \right) ^\beta,
\label{powereq2}
\end{align}
where $\tilde{\alpha} = \alpha \lambda^{\beta-1}$ is a dimensionless parameter.

\begin{figure}[tp]
\centering
\includegraphics[width= \columnwidth]{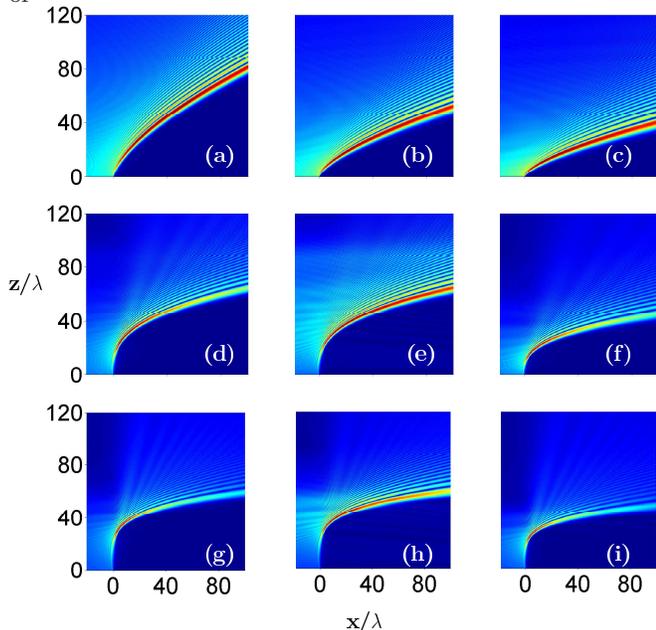} 
 \caption{Amplitude profile of accelerating beams following power-law trajectories for  $\gamma = 1/2 $ and
$\beta =3/2$, $\delta = 50 \lambda$  (upper panel) (a) $\tilde{\alpha} = 1$, (b) $\tilde{\alpha} = 2$, (c) $\tilde{\alpha} = 3$; $\beta =3$ (middle panel)(d) $\tilde{\alpha} = 1$, , $\delta = 50 \lambda$, (e) $\tilde{\alpha} = 1$, $\delta = 500 \lambda$, (f) $\tilde{\alpha} = 3$, $\delta = 50 \lambda$ ; $\beta =5$ (lower panel) (g) $\tilde{\alpha} = 1$, $\delta = 50 \lambda$ , (h) $\tilde{\alpha} = 1$, $\delta = 500 \lambda$, (i) $\tilde{\alpha} = 3$, $\delta = 50 \lambda$.
The white/black dashed line represents the analytical trajectory of the beam.  
\label{fig3}}
\end{figure}

The upper panel of Fig.~\ref{fig3} depicts the amplitude profile of a power law beam trajectory with exponent $\beta=3/2$ for different values of the scaling parameter $\tilde\alpha$ that affects the acceleration of the trajectory. As one can see, the beam amplitude is quite uniform along the caustic irrespectively of the value of $\tilde\alpha$. 
In the middle panel of Fig.~\ref{fig3} beams with cubic trajectories ($\beta=3$) are shown. In Fig.~\ref{fig3}(e) we modify the amplitude function as compared to Fig.~\ref{fig3}(d) resulting to a more intense caustic for larger values of $z$. In Fig~\ref{fig3}(f) we change the scaling factor $\tilde\alpha$ resulting to increased transverse acceleration. 
In the bottom panel of Fig.~\ref{fig3} accelerating trajectories with a larger exponent $\beta=5$ are presented. The presence of a ``knee'' in this case makes the intensity along the caustic trajectory less uniform and more intense around the ``knee''. This happens due to the large amount of rays that contribute to the caustic around the ``knee''. In Fig.~\ref{fig3}(h) we modify the initial amplitude profile as compared to Fig.~\ref{fig3}(g) resulting to a more uniform profile around the caustic after the ``knee''. In Fig.~\ref{fig3}(i) we change $\tilde\alpha$ and thus the acceleration is increased.

Finally, we apply the presented methodology for generating beams that follow exponential trajectories having the form
\begin{align} \label{expeq}
x_c = f(z_c) = x_0 e^{\beta (z_c-z_0)}.
\end{align}
We note that $x_c(z=0)=x_0e^{-\beta z_0}=\alpha$ and thus we should initially excite rays with $\xi<\alpha$. We select an initial amplitude profile  $A(\xi) = (\alpha + \delta - \xi)^{-\gamma} $ for $\xi \leq \alpha$ and $A(\xi) = 0$ for $\xi > \alpha$. 
Following the relevant calculations we compute the derivative of the initial phase 
\begin{align}
\frac{d \phi}{d \xi} = \frac{\alpha \beta k \exp { \left(1 + W\left(\zeta\right)\right)}}{\sqrt{1+ \alpha ^2 \beta^2 \exp{\left( 2 \left( 1+W\left( \zeta \right) \right) \right)}}} ,
\end{align}
where $W$ is the Lambert $W$ function, $\zeta=-\xi/(\alpha e)$, and $e$ is the Euler's number. Integrating the above equation, we find the initial phase profile:
\begin{align}
\phi(\xi) = -g(\xi)+g(\alpha)
\end{align}
where
$
g(\xi) = (k/\beta) \left[\tanh ^{-1}f(\xi)
 + f(\xi) W \left( \zeta \right)\right],
$
 and
$
f(\xi) = \sqrt{
1+(\beta^2\xi^2/W^2(\zeta))
}.
$

\begin{figure}[tp]
\centering
\includegraphics[width= \columnwidth]{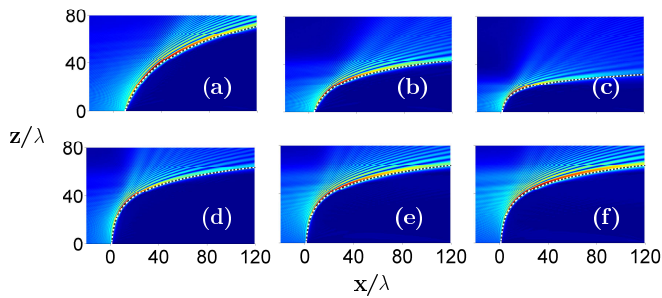} 
 \caption{Amplitude profile of accelerating beams following exponential caustic trajectories for  $x_0 = 20 \lambda$. In the upper panel $z_0 = x_0$, $\delta = x_0$ and (a) $\beta=0.7/ x_0$, (b) $\beta = 3/2 x_0$, (c) $\beta = 3/ x_0$. In the lower panel $z_0 = 2 x_0$, $\beta =3/2 x_0$ and (d) $\delta= x_0$, (e) $\delta = 5 x_0$, (c) $\delta = 10 x_0$. In all cases $\gamma =1/2$. The white/black dashed line represents the analytical trajectory of the beam. \label{fig4}}
\end{figure}
In the upper panel of Fig.~\ref{fig4} accelerating exponential trajectories with different exponents are shown. As the exponent increases, the acceleration increases and the beam bends faster. In the lower panel of Fig.~\ref{fig4} we use different amplitude truncations leading to a caustic trajectory that is more intense in the beginning the intermediate or the later stages, respectively. 

Finally, let us point out that in all the figures one can see that the numerically obtained beam trajectories are in agreement with the expected geometrical curves (dashed white-black lines) [given by Eqs.\eqref{circ_caust_eq}, \eqref{elip_caust_eq}, \eqref{powereq} and \eqref{expeq}].

In conclusion, in this letter we proposed a real-space method for generating nonparaxial accelerating caustics with arbitrary trajectories. We utilized the method for finding, in closed-form, the required initial phase for generating accelerating beams following circular, elliptical, and 
for the first time general power-law and exponential trajectories. 
\rev{We would like to point out that similar methods can be applied to different nonparaxial settings, for example, for generating  vector caustics~\cite{dribe-ol2014} or for designing different classes of nonparaxial trajectories~\cite{bride-ol2013}.}


Supported by the action ``ARISTEIA'' (grant no. 1261) in the context of the Operational Programme "Education and Lifelong Learning" that is co-funded by the European Social Fund and National Resources and by the 
Research Project ANEMOS co-financed by the European
Union (European Social Fund - ESF) and Greek national funds
through the Operational Program ``Education and Lifelong
Learning'' of the National Strategic Reference Framework
(NSRF) - Research Funding Program: Thales.



\begin{thebibliography}{10}
\newcommand{\enquote}[1]{``#1''}

\bibitem{sivil-ol2007}
G.~A. Siviloglou and D.~N. Christodoulides, \enquote{Accelerating finite energy
  {A}iry beams,} Opt. Lett. \textbf{32}, 979--981 (2007).

\bibitem{sivil-prl2007}
G.~A. Siviloglou, J.~Broky, A.~Dogariu, and D.~N. Christodoulides,
  \enquote{Observation of accelerating {A}iry beams,} Phys. Rev. Lett.
  \textbf{99}, 213901 (2007).

\bibitem{hu-springer2012}
Y.~Hu, G.~A. Siviloglou, P.~Zhang, N.~K. Efremidis, D.~N. Christodoulides, and
  Z.~Chen, \enquote{Self-accelerating airy beams: Generation, control, and
  applications,} in \enquote{Nonlinear Photonics and Novel Optical Phenomena,}
  , vol. 170 of \emph{Springer Series in Optical Sciences}, Z.~Chen and
  R.~Morandotti, eds. (Springer New York, 2012), pp. 1--46.

\bibitem{berry-ajp1979}
M.~V. Berry and N.~L. Balazs, \enquote{Nonspreading wave packets,} Am. J. of
  Phys. \textbf{47}, 264--267 (1979).

\bibitem{green-prl2011}
E.~Greenfield, M.~Segev, W.~Walasik, and O.~Raz, \enquote{Accelerating light
  beams along arbitrary convex trajectories,} Phys. Rev. Lett. \textbf{106},
  213902 (2011).

\bibitem{chrem-ol2011}
I.~Chremmos, N.~K. Efremidis, and D.~N. Christodoulides,
  \enquote{Pre-engineered abruptly autofocusing beams,} Opt. Lett. \textbf{36},
  1890--1892 (2011).

\bibitem{froeh-oe2011}
L.~Froehly, F.~Courvoisier, A.~Mathis, M.~Jacquot, L.~Furfaro, R.~Giust, P.~A.
  Lacourt, and J.~M. Dudley, \enquote{Arbitrary accelerating micron-scale
  caustic beams in two and three dimensions,} Opt. Express \textbf{19},
  16455--16465 (2011).

\bibitem{polyn-science2009}
P.~Polynkin, M.~Kolesik, J.~V. Moloney, G.~A. Siviloglou, and D.~N.
  Christodoulides, \enquote{Curved plasma channel generation using ultraintense
  {A}iry beams,} Science \textbf{324}, 229--232 (2009).

\bibitem{baumg-np2008}
J.~Baumgartl, M.~Mazilu, and K.~Dholakia, \enquote{Optically mediated particle
  clearing using {A}iry wavepackets,} Nat. Photon. \textbf{2}, 675--678 (2008).

\bibitem{zhang-ol2011}
P.~Zhang, J.~Prakash, Z.~Zhang, M.~S. Mills, N.~K. Efremidis, D.~N.
  Christodoulides, and Z.~Chen, \enquote{Trapping and guiding microparticles
  with morphing autofocusing airy beams,} Opt. Lett. \textbf{36}, 2883--2885
  (2011).

\bibitem{mathi-apl2012}
A.~Mathis, F.~Courvoisier, L.~Froehly, L.~Furfaro, M.~Jacquot, P.~Lacourt, and
  J.~Dudley, \enquote{Micromachining along a curve: Femtosecond laser
  micromachining of curved profiles in diamond and silicon using accelerating
  beams,} Appl. Phys. Lett. \textbf{101} (2012).

\bibitem{salar-ol2010}
A.~Salandrino and D.~N. Christodoulides, \enquote{Airy plasmon: a
  nondiffracting surface wave,} Opt. Lett. \textbf{35}, 2082--2084 (2010).

\bibitem{zhang-ol2011-plasmon}
P.~Zhang, S.~Wang, Y.~Liu, X.~Yin, C.~Lu, Z.~Chen, and X.~Zhang,
  \enquote{Plasmonic {A}iry beams with dynamically controlled trajectories,}
  Opt. Lett. \textbf{36}, 3191--3193 (2011).

\bibitem{minov-prl2011}
A.~Minovich, A.~E. Klein, N.~Janunts, T.~Pertsch, D.~N. Neshev, and Y.~S.
  Kivshar, \enquote{Generation and near-field imaging of airy surface
  plasmons,} Phys. Rev. Lett. \textbf{107}, 116802 (2011).

\bibitem{efrem-ol2010}
N.~K. Efremidis and D.~N. Christodoulides, \enquote{Abruptly autofocusing
  waves,} Opt. Lett. \textbf{35}, 4045--4047 (2010).

\bibitem{papaz-ol2011}
D.~G. Papazoglou, N.~K. Efremidis, D.~N. Christodoulides, and S.~Tzortzakis,
  \enquote{Observation of abruptly autofocusing waves,} Opt. Lett. \textbf{36},
  1842--1844 (2011).

\bibitem{voloc-nature2013}
N.~Voloch-Bloch, Y.~Lereah, Y.~Lilach, A.~Gover, and A.~Arie,
  \enquote{Generation of electron airy beams,} Nature \textbf{494}, 331--335
  (2013).

\bibitem{efrem-pra2013}
N.~K. Efremidis, V.~Paltoglou, and W.~von Klitzing, \enquote{Accelerating and
  abruptly autofocusing matter waves,} Phys. Rev. A \textbf{87}, 043637 (2013).

\bibitem{chrem-ieee2013}
I.~Chremmos, G.~Fikioris, and N.~Efremidis, \enquote{Accelerating and
  abruptly-autofocusing beam waves in the fresnel zone of antenna arrays,}
  Antennas and Propagation, IEEE Transactions on \textbf{61}, 5048--5056
  (2013).

\bibitem{chrem-ol2012-bessel}
I.~D. Chremmos, Z.~Chen, D.~N. Christodoulides, and N.~K. Efremidis,
  \enquote{Bessel-like optical beams with arbitrary trajectories,} Opt. Lett.
  \textbf{37}, 5003--5005 (2012).

\bibitem{zhao-ol2013}
J.~Zhao, P.~Zhang, D.~Deng, J.~Liu, Y.~Gao, I.~D. Chremmos, N.~K. Efremidis,
  D.~N. Christodoulides, and Z.~Chen, \enquote{Observation of self-accelerating
  bessel-like optical beams along arbitrary trajectories,} Opt. Lett.
  \textbf{38}, 498--500 (2013).

\bibitem{courv-ol2012}
F.~Courvoisier, A.~Mathis, L.~Froehly, R.~Giust, L.~Furfaro, P.~A. Lacourt,
  M.~Jacquot, and J.~M. Dudley, \enquote{Sending femtosecond pulses in circles:
  highly nonparaxial accelerating beams,} Opt. Lett. \textbf{37}, 1736--1738
  (2012).

\bibitem{kamin-prl2012}
I.~Kaminer, R.~Bekenstein, J.~Nemirovsky, and M.~Segev, \enquote{Nondiffracting
  accelerating wave packets of maxwell's equations,} Phys. Rev. Lett.
  \textbf{108}, 163901 (2012).

\bibitem{zhang-prl2012}
P.~Zhang, Y.~Hu, T.~Li, D.~Cannan, X.~Yin, R.~Morandotti, Z.~Chen, and
  X.~Zhang, \enquote{Nonparaxial {M}athieu and {W}eber accelerating beams,}
  Phys. Rev. Lett. \textbf{109}, 193901 (2012).

\bibitem{aleah-prl2012}
P.~Aleahmad, M.-A. Miri, M.~S. Mills, I.~Kaminer, M.~Segev, and D.~N.
  Christodoulides, \enquote{Fully vectorial accelerating diffraction-free
  {H}elmholtz beams,} Phys. Rev. Lett. \textbf{109}, 203902 (2012).

\bibitem{bandr-njp2013}
M.~A. Bandres and B.~M. Rodr\'iguez-Lara, \enquote{Nondiffracting accelerating
  waves: Weber waves and parabolic momentum,} New J. Phys. \textbf{15}, 013054
  (2013).

\bibitem{mathi-ol2013}
A.~Mathis, F.~Courvoisier, R.~Giust, L.~Furfaro, M.~Jacquot, L.~Froehly, and
  J.~M. Dudley, \enquote{Arbitrary nonparaxial accelerating periodic beams and
  spherical shaping of light,} Opt. Lett. \textbf{38}, 2218--2220 (2013).

\bibitem{epste-prl2014}
I.~Epstein and A.~Arie, \enquote{Arbitrary bending plasmonic light waves,}
  Phys. Rev. Lett. \textbf{112}, 023903 (2014).

\bibitem{chrem-pra2013}
I.~D. Chremmos and N.~K. Efremidis, \enquote{Nonparaxial accelerating
  bessel-like beams,} Phys. Rev. A \textbf{88}, 063816 (2013).

\bibitem{dribe-ol2014}
R.~Driben, V.~V. Konotop, and T.~Meier, \enquote{Coupled airy breathers,} Opt. Lett. \textbf{39}, 5523--5526 (2014).

\bibitem{bride-ol2013}
R.~Driben, Y.~Hu, Z.~Chen, B.~A. Malomed, and R.~Morandotti, \enquote{Inversion
  and tight focusing of airy pulses under the action of third-order
  dispersion,} Opt. Lett. \textbf{38}, 2499--2501 (2013).

\end{thebibliography}

\newcommand{\noopsort[1]}{} \newcommand{\singleletter}[1]{#1}

\cleardoublepage

\noindent\textbf{References with titles}

\end{document}